\documentclass[prb,twocolumn,amsmath,amssymb]{revtex4}
\usepackage{xcolor}

\usepackage{wasysym}
\usepackage{graphicx}
\usepackage{dcolumn}
\usepackage{bm}
\usepackage{epstopdf}

\def\E{\rm E}

\def\S{\rm S}
\def\be{\begin{equation}}
\def\ee{\end{equation}}
\def\bd{\begin{displaymath}}
\def\ed{\end{displaymath}}
\def\-{\phantom{-}}

\begin{document}

\title{Evolution of the thermodynamic properties and inelastic neutron scattering intensities for spin-1/2 antiferromagnetic quantum rings}

\author{Jason T. Haraldsen\footnote{email: j.t.haraldsen@unf.edu}}
\affiliation{Department of Physics, University of North Florida, Jacksonville, FL 32224}

\date{\today}

\begin{abstract}

This study examines the increasing complexity in the magnetic properties of small $n$ = 3, 4, 5, 6 spin-1/2 quantum rings. Using an exact diagonalization of the isotropic Heisenberg Hamiltonian with nearest and next-nearest neighbor interactions, the energy eigenstates, magnetic specific heat capacity, magnetic susceptibility, and inelastic neutron scattering structure factors are determined for variable next-nearest neighbor interactions. Here, it is shown that the presence of a complex spin-mixing, multiple ground states, and non-zero ground states greatly complicate the spin Hamiltonian. Overall, the energy eigenstates and structure factor intensities are presented in closed form, while the thermodynamic properties detail the effect of a crossing interaction in the rings. The goal of this work is to provide insight into the evolution of the magnetic properties and spin excitations within these systems.

\end{abstract}

\maketitle

\section{Introduction}

Nanomagnetism continues to be a highly-active research area in condensed matter physics due to the potential for spintronic devices through the coupling of electronic and magnetic degrees of freedom\cite{whit:10,Kah93,gatt:06,klemm:08,tasi:04,barc:05,chab:02,dunb:07,shat:09,dua:16,zhe:16,gos:15,gao:14}. The interactions in nanomagnetic systems range from long-range interactions that produce spin-waves and skyrmionic orders to short-range magnetic cluster that exhibit discrete excitations. Recently, the desire to control magnetic ground states for the propose of technological advancement in areas like quantum computation and magnetic storage\cite{Nie00,tyag:09} has produced a focus on magnetic nanoparticles, molecular-based magnets, and generalized spin clusters because of their ability to be manipulated on the nanoscale.

Many molecular magnets can be generalized as mainly isolated magnetic interactions either as small magnetic clusters, molecular solids or due to weaken orbital interactions. Magnetic clusters are typically considered to be a grouping of magnetic ions, in various geometric configurations, that are isolated from long-range magnetic interactions through non-magnetic ligands\cite{whit:10,Kah93,gatt:06,klemm:08,tasi:04,barc:05,chab:02,dunb:07,shat:09,dua:16,zhe:16,gos:15,gao:14}. These include Mn$_{12}$ and Mn$_{84}$ or Fe$_{8}$\cite{barc:05,chab:02,tasi:04}. 

However, other molecular magnet and spin cluster systems consist of isolated magnetic interactions even in the presence of a long-range lattice structures, where there is a breakdown of orbital overlap and interactions that weaken or eliminate long-magnetic order and produce localized magnetic clusters with discrete or weakly dispersive excitations\cite{kawa:16}. Excitations like this have been observed in certain pyrochlore systems\cite{lee:00,lee:02,tomi:13} as well as the vanadium dimer of VOHPO$_4\cdot$0.5H$_2$O\cite{ten:97,koo:02}. 

More recently, there has been an increased interest in understanding magnetic atom and cluster interactions either substituted in or adsorbed on two-dimensional materials like MoS$_2$ and graphene\cite{muru:05,eel:13,ndi:06,cab:10,cer:16}. In these cases, researchers are looking to understand how the magnetic states enhance the electronic properties of Dirac material\cite{wehl:14}. However, the first step to gaining an appreciation for these complex systems is by examining the magnetic interactions and the bulk and microscopic properties produced by various clusters.

\begin{figure}
\includegraphics[width=3.25in]{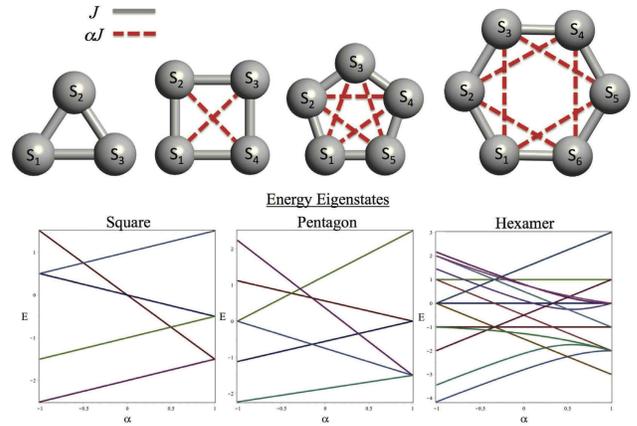}
\vskip -0.2cm
\caption{(color online) The structures for the spin-1/2 triangle, square, pentagon, and hexagon, and the energy levels as a function of $\alpha J$ for the latter three.}
\label{structures}
\end{figure}

The simplest magnetic cluster is the magnetic dimer (a two body spin cluster). However, this system has been exhaustively investigated by a number of studies\cite{houc:15,Wha03,furr:13}. In 2005, the excitations of dimers, trimers, and tetramers were examined in detail\cite{hara:05}. Typically, the magnetic properties and inelastic structures factors for different geometries are determined on a case-by-case basis or numerically on larger systems\cite{ten:97,hara:09,borr:99,henn:97}, which does not provide any insight into the scalability and growth of these systems.


\begin{table*}
\caption{Energy Levels and Excitations for the Spin-1/2 Triangle and Square}
\vskip 0.5cm
\begin{ruledtabular}
\begin{tabular}{lccc}
$|{\rm S}_{tot}\big>^{deg.}$ & Energy Level & Excitations & Structure Factor\\
 & & ($\alpha = 0$) & Functional Form \\
\colrule
\hline
Triangle \\
\hline
&   \\
$|\frac{3}{2}\big>$ &  $\frac{3}{4}J$ & $\frac{3}{2}J$ & $3 + \cos(\vec q \cdot \vec r_{13})- 2\cos(\vec q \cdot \vec r_{12})- 2\cos(\vec q \cdot \vec r_{23})$\\
&&&and\\
&&& $1 - \cos(\vec q \cdot \vec r_{13})$\\
\\ \hline
&   \\
$|\frac{1}{2}\big>^2$ &  -$\frac{3}{4}J$ & Ground State & \\ 
\\ \hline \hline
Square \\
\hline
&   \\
$|2\big>$ &  $\frac{J}{2}(\alpha$ + $2)$ \\ 
\\
\hline
\\
$|1\big>_{II}^2$ &  -$\frac{1}{2}\alpha J$ & $2J$ & $2 - \cos(\vec q \cdot \vec r_{12})- \cos(\vec q \cdot \vec r_{13})+ \cos(\vec q \cdot \vec r_{14})$\\&&&$+ \cos(\vec q \cdot \vec r_{23})- \cos(\vec q \cdot \vec r_{24})- \cos(\vec q \cdot \vec r_{34})$\\
&&&and\\
&&& $2 + \cos(\vec q \cdot \vec r_{12})- \cos(\vec q \cdot \vec r_{13})- \cos(\vec q \cdot \vec r_{14})$\\&&&$- \cos(\vec q \cdot \vec r_{23})- \cos(\vec q \cdot \vec r_{24})+ \cos(\vec q \cdot \vec r_{34})$\\
\\ 
$|1\big>_{I}$ &   $\frac{J}{2}(\alpha$-$2)$ & $J$ & $2 - \cos(\vec q \cdot \vec r_{12})+ \cos(\vec q \cdot \vec r_{13})- \cos(\vec q \cdot \vec r_{14})$\\&&&$- \cos(\vec q \cdot \vec r_{23})+ \cos(\vec q \cdot \vec r_{24})- \cos(\vec q \cdot \vec r_{34})$\\
\\
\hline
\\
$|0\big>_{II}$ &  -$\frac{3}{2}\alpha J$ & Ground State & 0 \\
 & & ($\alpha >$ 1) \\
$|0\big>_{I}$ & $\frac{J}{2}(\alpha $ - $4)$ & Ground State \\
 & & ($\alpha <$ 1)\\    
\label{t-t-values}
\end{tabular}
\end{ruledtabular}
\end{table*}

Considering that many nanoparticle and molecular magnet systems consist of atomistic rings or ring-like clusters, the ability to gain understanding of the evolution of quantum rings is relevant to number of materials and research areas in condensed matter physics. Therefore, in this article, the energy eigenstates, magnetic properties, and inelastic neutron scattering structure factor for small spin-1/2 quantum rings are determined through the exact diagonalization of the isotropic Heisenberg Hamiltonian. The evolution of the these properties is shown for rings consisting of $n$ magnetic spins ($n$ = 3, 4, 5, and 6) with nearest- and next-nearest neighbor interactions (if applicable). The purpose of this study is to provide the general understanding of the magnetic properties for these structures to help guide the experimental characterization of cluster interactions in various materials like molecular magnets and two-dimensional materials. Furthermore, the calculation of the inelastic neutron scattering structure factors for multiple subgeometries is provided.

\section{Energy Eigenstates}

To determine the spin eigenvalues and eigenstates for the spin-1/2 quantum systems, we first consider the isotropic Heisenberg Hamiltonian,
\be
{\cal H} = \sum_{<ij>}  {\rm J}_{ij}\;  \vec{\rm S}_{i}\cdot
\vec{\rm S}_{j},
\label{magH}
\ee
where the superexchange constants $\{ {\rm J}_{ij}\} $ are positive for antiferromagnetic interactions and negative for ferromagnetic ones, and $\vec {\rm S}_i$ is the quantum spin operator for a spin-1/2 ion at site $i$\cite{hara:05,Kah93}. 

\begin{table*}
\caption{Energy Levels and Excitations for the Spin-1/2 Pentagon}
\vskip 0.5cm
\begin{ruledtabular}
\begin{tabular}{lccc}
$|{\rm S}_{tot}\big>^{deg.}$ & Energy Level & Excitations & Structure Factor\\
 & & ($\alpha = 0$) & Functional Form \\
\colrule
\hline
Pentagon\footnote{$a_1$=$\frac{1}{2}(\sqrt{5}+3)$, $a_2$=$\frac{1}{2}(\sqrt{5}+1)$, $a_3$=$\frac{1}{2}(\sqrt{5}+2$), $a_4$=$\frac{1}{2}(\sqrt{5}-3)$, $a_5$=$\frac{1}{2}(1-\sqrt{5})$,  $a_6$=$\frac{1}{2}(2-\sqrt{5})$, and $a_7$=$\frac{1}{2}(3\sqrt{5}+7)$} \\
\hline
&   \\
$|\frac{5}{2}\big>_{I}$ &  $\frac{5J}{4}(\alpha + 1)$ & Ground State \\
& & ($\alpha < \frac{\sqrt{5}+4}{\sqrt{5}-4}$) \\
\\
\hline
\\
$|\frac{3}{2}\big>^2_{II}$ &  -$\frac{J}{4}\sqrt{5}(\alpha - 1)$&$\frac{3J}{4} \Big( (1- \alpha) \sqrt {5}$&$1+a_1^2+2a_2^2 + \cos(\vec q \cdot \vec r_{12}) -a_1\cos(\vec q \cdot \vec r_{13})+2a_2\cos(\vec q \cdot \vec r_{14}) $ \\ &&~~~~~~~~~~~~~~$+\alpha+1 \Big) $& $-a_1\cos(\vec q \cdot \vec r_{15}) -a_1\cos(\vec q \cdot \vec r_{23}) +2a_2\cos(\vec q \cdot \vec r_{24}) -a_1\cos(\vec q \cdot \vec r_{25}) $ \\ &&&$-2a_1a_2\cos(\vec q \cdot \vec r_{34})+a_1^2\cos(\vec q \cdot \vec r_{35}) -2a_1a_2\cos(\vec q \cdot \vec r_{45})$ \\ 
&&&and\\
&&& $a_1^2+a_7^2 +8a_3^2 - a_1a_7 \cos(\vec q \cdot \vec r_{12}) - a_1a_7\cos(\vec q \cdot \vec r_{13})+ a_7^2\cos(\vec q \cdot \vec r_{14}) $ \\ &&& $-4a_3a_7\cos(\vec q \cdot \vec r_{15}) + a_1^2\cos(\vec q \cdot \vec r_{23}) - a_1a_7\cos(\vec q \cdot \vec r_{24})+4a_1a_3\cos(\vec q \cdot \vec r_{25})  $ \\ &&&$-a_1a_7\cos(\vec q \cdot \vec r_{34})+4a_1a_3\cos(\vec q \cdot \vec r_{35}) -4a_3a_7\cos(\vec q \cdot \vec r_{45})$ \\ 
&&&and\\
&&& $1+a_4^2+2a_5^2 + a_4\cos(\vec q \cdot \vec r_{12}) +2a_4a_5\cos(\vec q \cdot \vec r_{13})+a_4\cos(\vec q \cdot \vec r_{14}) $ \\ &&& $+a_4^2\cos(\vec q \cdot \vec r_{15}) +2a_5\cos(\vec q \cdot \vec r_{23}) +\cos(\vec q \cdot \vec r_{24}) +a_4\cos(\vec q \cdot \vec r_{25}) $ \\ &&&$+2a_5\cos(\vec q \cdot \vec r_{34})+2a_4a_5\cos(\vec q \cdot \vec r_{35}) +a_4\cos(\vec q \cdot \vec r_{45})$ \\ 
\\
$|\frac{3}{2}\big>^2_{I}$ &   $\frac{J}{4}\sqrt{5}(\alpha - 1)$ & $\frac{J}{4}\Big( (1-\alpha) \sqrt {5}$& $1+a_7^2 +18a_1^2 + \cos(\vec q \cdot \vec r_{12}) + a_7\cos(\vec q \cdot \vec r_{13})-6a_1\cos(\vec q \cdot \vec r_{14}) $ \\ &&~~~~~~~~~~~$+3(\alpha-1) \Big)$& $+a_7\cos(\vec q \cdot \vec r_{15}) +a_7\cos(\vec q \cdot \vec r_{23}) -6a_1\cos(\vec q \cdot \vec r_{24}) +a_7\cos(\vec q \cdot \vec r_{25}) $ \\ &&&$-6a_7a_1\cos(\vec q \cdot \vec r_{34})+a_7^2\cos(\vec q \cdot \vec r_{35}) -6a_7a_1\cos(\vec q \cdot \vec r_{45})$ \\
&&&and\\
&&& $1+13a_5^2 + 18a_6^2 + (5a_4+1)^2 - 36a_5a_6\cos(\vec q \cdot \vec r_{12}) $ \\ &&& $ +12(5a_4+1)a_6\cos(\vec q \cdot \vec r_{13})  +12a_6\cos(\vec q \cdot \vec r_{14}) +24a_6a_5\cos(\vec q \cdot \vec r_{15}) $ \\ &&& $-6(5a_4+1)a_5\cos(\vec q \cdot \vec r_{23}) - 6a_5\cos(\vec q \cdot \vec r_{24}) -12a_5^2\cos(\vec q \cdot \vec r_{25}) $ \\ &&&$+2(5a_4+1)\cos(\vec q \cdot \vec r_{34})+4(5a_4+1)a_5\cos(\vec q \cdot \vec r_{35}) +4a_5\cos(\vec q \cdot \vec r_{45})$ \\
&&&and\\
&& & $50-3a_2(2a_3+5)\cos(\vec q \cdot \vec r_{12/45}) +12\cos(\vec q \cdot \vec r_{13/14})-18\cos(\vec q \cdot \vec r_{14/24}) $ \\ &&& $-6a_5\cos(\vec q \cdot \vec r_{15/34}) + (5a_5-a_1)(2\cos(\vec q \cdot \vec r_{23/15}) -3\cos(\vec q \cdot \vec r_{24/25})) $ \\ &&&$+(5a_4-a_2)\cos(\vec q \cdot \vec r_{25/35}) -12a_5^2\cos(\vec q \cdot \vec r_{34/12})-4a_5^3\cos(\vec q \cdot \vec r_{35/13}) $\\ &&&$+6a_5^3\cos(\vec q \cdot \vec r_{45/23})$ \\
\\
\hline
\\
$|\frac{1}{2}\big>^2_{III}$ &  -$\frac{J}{4}\big((2\sqrt{5} + 3)\alpha + (3 - 2\sqrt{5})\big)$ & Ground State & $2 + \cos(\vec q \cdot \vec r_{12/13}) - \cos(\vec q \cdot \vec r_{13/14}) - \cos(\vec q \cdot \vec r_{15/15})$\\&&($\alpha >$ 1)&$- \cos(\vec q \cdot \vec r_{23/34})- \cos(\vec q \cdot \vec r_{25/35}) + \cos(\vec q \cdot \vec r_{35/45})$\\ &&& and \\
&&$\sqrt{5}(1-\alpha)J$&$ a_1^2+ a_2^2+ 4a_3^2-4a_2a_3\cos(\vec q \cdot \vec r_{15}) + 2a_1a_2\cos(\vec q \cdot \vec r_{13}) - 4a_1a_3\cos(\vec q \cdot \vec r_{35}) $\\
&&& and\\
&&&$ a_4^2+a_5^2+4a_6^2 - 4a_5a_6\cos(\vec q \cdot \vec r_{15})-2a_4a_5\cos(\vec q \cdot \vec r_{13}) + 4a_4a_6\cos(\vec q \cdot \vec r_{35}) $\\\\
$|\frac{1}{2}\big>_{II}$ &  -$\frac{3J}{4}(\alpha + 1)$& $\frac{\sqrt{5}}{2}(1-\alpha)J$ & $a_1^2+a_2^2 - a_2^2\cos(\vec q \cdot \vec r_{21})- a_1a_2\cos(\vec q \cdot \vec r_{13})+ a_1a_2\cos(\vec q \cdot \vec r_{15})$\\&&&$+ a_1a_2\cos(\vec q \cdot \vec r_{23})- a_1a_2\cos(\vec q \cdot \vec r_{25})- a_1^2\cos(\vec q \cdot \vec r_{35})$\\

&&&and\\
&&& $1+a_5^2 - \cos(\vec q \cdot \vec r_{13}) + a_5\cos(\vec q \cdot \vec r_{14})- a_5\cos(\vec q \cdot \vec r_{15})$\\&&&$- a_5\cos(\vec q \cdot \vec r_{34})+ a_5\cos(\vec q \cdot \vec r_{35})- a_5^2\cos(\vec q \cdot \vec r_{45})$\\

$|\frac{1}{2}\big>^2_{I}$ &   $\frac{J}{4}\big((2\sqrt{5} - 3)\alpha - (3 + 2\sqrt{5})\big)$ & Ground State \\  
 &   & ($ \frac{\sqrt{5}+4}{\sqrt{5}-4} < \alpha <$ 1) \\
\label{p-values}
\end{tabular}
\end{ruledtabular}
\end{table*}

\begin{table*}
\caption{Energy Levels and First Excitation for the Spin-1/2 Hexagon}
\vskip 0.5cm
\begin{ruledtabular}
\begin{tabular}{lccc}
$|{\rm S}_{tot}\big>^{deg.}$ & Energy Level & Excitation & Structure Factor \\
 &  & ($\alpha=0$) & Functional Form \\
\colrule
\hline
&   \\
$|3\big>$ & $\frac{3J}{2}(\alpha$+$1)$ \\ 
\\ \hline
&   \\
$|2\big>^2_{III}$ & $J$ \\ 
$|2\big>^2_{II}$ & $0$ \\ 
$|2\big>_{I}$ & $\frac{J}{2}(3\alpha$-$1)$ \\ 
\\ \hline
&   \\
$|1\big>^2_{VI}$ & -$\frac{J}{4}\big(3\alpha+1-\sqrt{9\alpha^2 - 10\alpha + 17}\big)$ &$\frac{J}{4}(3+\sqrt {17}+2\sqrt {13})$ &$2 - \cos(\vec q \cdot \vec r_{13/15})- \cos(\vec q \cdot \vec r_{14/14})+ \cos(\vec q \cdot \vec r_{16/12})$\\&&&$+ \cos(\vec q \cdot \vec r_{34/45})- \cos(\vec q \cdot \vec r_{36/24})- \cos(\vec q \cdot \vec r_{46/25})$\\ \\
$|1\big>_{V}$ & $\frac{J}{2}(1$-$3\alpha)$ &$\frac{J}{2}(3+\sqrt {13})$& 0\\ \\
$|1\big>_{IV}$ & -$\frac{J}{2}\big(2-\sqrt{9\alpha^2 - 10\alpha + 5}\big)$  &$\frac{J}{2}(\sqrt{13}+\sqrt{5})$& $3 + \sum_{ij} (-1)^{i+j} \cos(\vec q \cdot \vec r_{ij})$  \\ \\
$|1\big>^2_{III}$ & -$J$ &$\frac{J}{2}\sqrt {13}$& $6 + \cos(\vec q \cdot \vec r_{12/13}) - 2\cos(\vec q \cdot \vec r_{13/12}) + \cos(\vec q \cdot \vec r_{14}) $ \\ &&&$ +\cos(\vec q \cdot \vec r_{15/16}) - 2\cos(\vec q \cdot \vec r_{16/15}) -2\cos(\vec q \cdot \vec r_{23}) $ \\&&&$ +\cos(\vec q \cdot \vec r_{24/34})+\cos(\vec q \cdot \vec r_{25/36})-2\cos(\vec q \cdot \vec r_{26}) $ \\ &&& $-2\cos(\vec q \cdot \vec r_{34/24})-2\cos(\vec q \cdot \vec r_{35})+4\cos(\vec q \cdot \vec r_{36/25}) $ \\ &&& $ +\cos(\vec q \cdot \vec r_{45/46})-2\cos(\vec q \cdot \vec r_{46/45})-2\cos(\vec q \cdot \vec r_{56})$ \\ \\
$|1\big>^2_{II}$ & -$\frac{J}{4}\big(3\alpha + 1+\sqrt{9\alpha^2 - 10\alpha + 17}\big)$ &$\frac{J}{4}(3-\sqrt {17}+2\sqrt {13})$ &$2 - \cos(\vec q \cdot \vec r_{13/15})- \cos(\vec q \cdot \vec r_{14/14})+ \cos(\vec q \cdot \vec r_{16/12})$\\&&&$+ \cos(\vec q \cdot \vec r_{34/45})- \cos(\vec q \cdot \vec r_{36/24})- \cos(\vec q \cdot \vec r_{46/25})$\\ \\ 
$|1\big>_{I}$ & -$\frac{J}{2}\big(2+\sqrt{9\alpha^2 - 10\alpha + 5}\big)$ & $\frac{J}{2}(\sqrt{13}-\sqrt{5})$ & $3 + \sum_{ij} (-1)^{i+j} \cos(\vec q \cdot \vec r_{ij})$\\ 
\\ \hline
&   \\
$|0\big>_{IV}$ & -$\frac{J}{2}\big(2-\sqrt{9\alpha^2 - 18\alpha + 13}\big)$ &$J\sqrt {13}$&0 \\ 
$|0\big>^2_{III}$ & -$\frac{J}{2}(3\alpha$+$1)$ &$\frac{J}{2}(1+\sqrt {13})$&0 \\ 
$|0\big>_{II}$ & -$\frac{3J}{2}(\alpha $+$1)$ & Ground State &0\\
& & & ($\alpha >  \frac{1}{2}$) \\ 
$|0\big>_{I}$ & -$\frac{J}{2}\big(2+\sqrt{9\alpha^2 - 18\alpha +13}\big)$ & Ground State \\ 
& & &($\alpha < \frac{1}{2}$) \\ 
\label{h-values}
\end{tabular}
\end{ruledtabular}
\end{table*}

For a each configuration shown in Fig. \ref{structures}, the nearest and next-nearest neighbor spin interactions are considered. Therefore, the spin-spin exchange Hamiltonians are 
\be 
\begin{array}{c}
{\cal H}_{\triangle} = {\rm J}\, \big( \vec{\rm S}_{1}\cdot\vec{\rm S}_{2} + \vec{\rm S}_{2}\cdot\vec{\rm S}_{3} +\vec{\rm S}_{1}\cdot\vec{\rm S}_{3} \big),\\
 {\cal H}_{\square} = {\rm J}\, \Big( \vec{\rm S}_{1}\cdot\vec{\rm S}_{2} + \vec{\rm S}_{2}\cdot\vec{\rm S}_{3} +\vec{\rm S}_{3}\cdot\vec{\rm S}_{4} +\vec{\rm S}_{1}\cdot\vec{\rm S}_{4}\\ +\alpha \big( \vec{\rm S}_{1}\cdot\vec{\rm S}_{3} + \vec{\rm S}_{2}\cdot\vec{\rm S}_{4}\big) \Big),\\
 {\cal H}_{\pentagon} = {\rm J}\, \Big( \vec{\rm S}_{1}\cdot\vec{\rm S}_{2} + \vec{\rm S}_{2}\cdot\vec{\rm S}_{3} +\vec{\rm S}_{3}\cdot\vec{\rm S}_{4} +\vec{\rm S}_{4}\cdot\vec{\rm S}_{5} +\vec{\rm S}_{1}\cdot\vec{\rm S}_{5} \\ 
 + \alpha \big( \vec{\rm S}_{1}\cdot\vec{\rm S}_{3} +  \vec{\rm S}_{1}\cdot\vec{\rm S}_{4} +  \vec{\rm S}_{2}\cdot\vec{\rm S}_{4} +  \vec{\rm S}_{2}\cdot\vec{\rm S}_{5} + \vec{\rm S}_{3}\cdot\vec{\rm S}_{5} \big) \Big),
\\
{\rm and}
\\
{\cal H}_{\hexagon} = {\rm J}\, \Big( \vec{\rm S}_{1}\cdot\vec{\rm S}_{2} + \vec{\rm S}_{2}\cdot\vec{\rm S}_{3} +\vec{\rm S}_{3}\cdot\vec{\rm S}_{4} +\vec{\rm S}_{4}\cdot\vec{\rm S}_{5} +\vec{\rm S}_{5}\cdot\vec{\rm S}_{6} \\
 + \vec{\rm S}_{1}\cdot\vec{\rm S}_{6} + \alpha \big( \vec{\rm S}_{1}\cdot\vec{\rm S}_{3} + \vec{\rm S}_{1}\cdot\vec{\rm S}_{5} + \vec{\rm S}_{2}\cdot\vec{\rm S}_{4} +  \vec{\rm S}_{2}\cdot\vec{\rm S}_{6} \\
 + \vec{\rm S}_{3}\cdot\vec{\rm S}_{5} \big) \Big).
\label{H} 
\end{array}
\ee

From this representation, we can look at the evolution of the spin decomposition
\be
\begin{array}{l}
\frac{1}{2} \otimes \frac{1}{2} \otimes \frac{1}{2} = \frac{1}{2}^2 \oplus \frac{3}{2}, \\
\\
\frac{1}{2} \otimes \frac{1}{2} \otimes \frac{1}{2} \otimes \frac{1}{2} = 0^2 \oplus 1^3 \oplus 2, \\
\\
\frac{1}{2} \otimes \frac{1}{2} \otimes \frac{1}{2} \otimes \frac{1}{2} \otimes \frac{1}{2} = \frac{1}{2}^5 \oplus \frac{3}{2}^4 \oplus \frac{5}{2}, \\
\\
{\rm and}\\ \\
\frac{1}{2} \otimes \frac{1}{2} \otimes \frac{1}{2} \otimes \frac{1}{2} \otimes \frac{1}{2} \otimes \frac{1}{2} = 3 \oplus 2^5 \oplus 1^9 \oplus 0^5. \\
\label{spins}
\end{array}
\ee
Each $S_{tot}$ multiplet contains 2$S_{tot}+1$ magnetic states, where the geometries consist of 8, 16, 32, and 64 total magnetic states and many degeneracies given the isotropic magnetic Hamiltonian. Therefore, two of the systems (equilateral triangle and pentagon) have non-zero ground states and two have complex spin-mixing in their states (pentagon and hexagon).

While the triangle and square states are fairly straightforward, the spin mixing of the pentagon and hexagon states complicate the evaluation of the magnetic properties. If the interactions between all spins in a cluster are equal, then no spin mixing occurs (i.e. the equilateral triangle) and the energy for all configurations can be given as
\be
E = \frac{J}{2}\Big[S_{tot}(S_{tot}+1) - \sum_i^n S_i(S_i+1) \Big],
\ee
where $S_{tot}$ is the total spin state for the cluster, $S_i$ denotes the spin at each site, $n$ are the total number of spins in the cluster ($n$ = 3 for the equilateral triangle.) The energy levels for the S = 1/2 equilateral triangle are given in Table \ref{t-t-values}.

When interactions vary between magnetic moments then spin mixing can occur depending on the geometries and symmetries in the cluster\cite{hara:11}. This doesn't occur in the case of the square, because the $\alpha J$ interactions are on separate dimers and the system can be broken into two coupled dimers. Therefore, the total energy for the square can be given in closed form as
\be
\begin{array}{l}
E_{\square}=\frac{J}{2}\Big[S_{tot}(S_{tot}+1) + (\alpha-1)(S_{d_1}(S_{d_1}+1) \\+ S_{d_2}(S_{d_2}+1) - \alpha\sum_i^4 S_i(S_i+1) \Big],
\end{array}
\ee
where $S_{d_i}$ is the total spin state for the individual dimers and $\alpha$ is the tuning parameter for the cross interactions. The individual states and energy levels are given in Table \ref{t-t-values}.

The pentagon and hexagon are different, because they introduce spin mixing. In the pentagon and hexagon configurations, there is no reason to assume the cross interactions are equal to the outside interactions.

In the pentagon (shown in Fig. \ref{structures}), the cross-interactions don't form distinct individual subgeometries, which makes a closed form representation of the eigenvalues nontrivial. Therefore, through a diagonalization of the full spin matrix for the pentagon, it is possible to determine the energy levels for the S = 1/2 pentagon (shown in Table \ref{p-values}.) It should noted that this is not true for all pentamers. Keep in mind that the configuration of the pentamer is critical in the determination of spin mixing. If the pentamer is made from a trimer coupled to an individual dimer (as discussed in Ref. [\onlinecite{hara:11}]), then no spin mixing occurs and the energies are solvable in closed form. This is similar for the hexagon.

The spin hexagon (shown in Fig. \ref{structures}) produces spin mixing because the cross interaction cannot be removed or accounted for without mixing subgeometries. In Fig. \ref{structures}, it looks like you can break down the structure into individual trimers (1-3-5, 2-4-6). However, the absent next-next-nearest neighbor interactions produce dimers (1-4, 2-5, 3-6) that are coupled to the trimers. However, as with the pentamer, this is not true for all hexamers. In Ref. [\onlinecite{hara:09}], the spin hexamer was composed of two weakly coupled trimers that accounted for all cross interactions by making them equal. Therefore, the Hamiltonian was easily diagonalizable and solvable in closed form. However, in the case of the hexagon, the decomposition of the spin configuration leads to mixed subgeometries and therefore, produces considerable mixing of the spin states. This is demonstrated by the calculated energy levels in Table \ref{h-values}.

\section{Thermodynamic Properties}

Through a diagonalization of the magnetic Hamiltonian, the energy eigenstates and eigenvalues may be found.
Using the determined energy eigenvalues, it is possible to determine the thermodynamic properties of specific heat and magnetic susceptibility. Since these properties are useful in the bulk characterization of magnetic materials, the general representation is calculated below. Here, the specific heat is given by
\be
C =  k_B \beta^{\, 2}\, \frac{d^2 \! \ln (Z)}{d \beta^2} \ ,
\label{C}
\ee
where $Z = \sum_{i=1}^N\, e^{- \beta \E_i}$, $k_B$ is Boltzmann's constant, $\beta = 1/k_BT$, and $E_i$ is the $i^{th}$ energy eigenstate. Furthermore, the magnetic susceptibility is determined from
\be
\chi = \frac{\beta}{Z}
\sum_{i=1}^N  \, (M_z^2)_i\, e^{-\beta \E_i} .
\label{chi}
\ee
where $M_z = m g \mu_B$ where $m = \S_{tot}^z/\hbar $ is the integral or half-integral magnetic quantum number, and $g$ is the electron $g$-factor. For simplicity, $\hbar$ is set to 1. Furthermore, in these formulas, there is a sum over $i=1\dots N$ is over all $N$ independent energy eigenstates (including magnetic substates), the sum $\sum_{\E_i}$ is over energy levels only.

\begin{figure}
\includegraphics[width=3.75in]{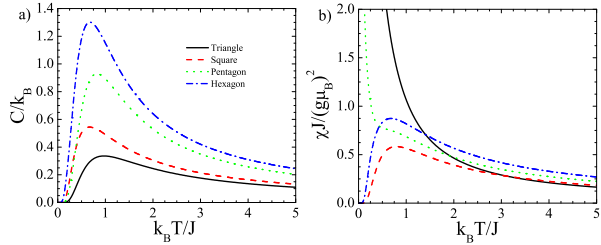}
\vskip -0.2cm
\caption{(color online) The calculated specific heat (a) and magnetic susceptibility (b) for the equilateral triangle (solid), square (dashed), pentagon (dotted), and hexagon (dash-dotted) as function of temperature for $\alpha$ = 0 (no cross interactions).}
\label{c-chi}
\end{figure}

\begin{figure}
\includegraphics[width=2.75in]{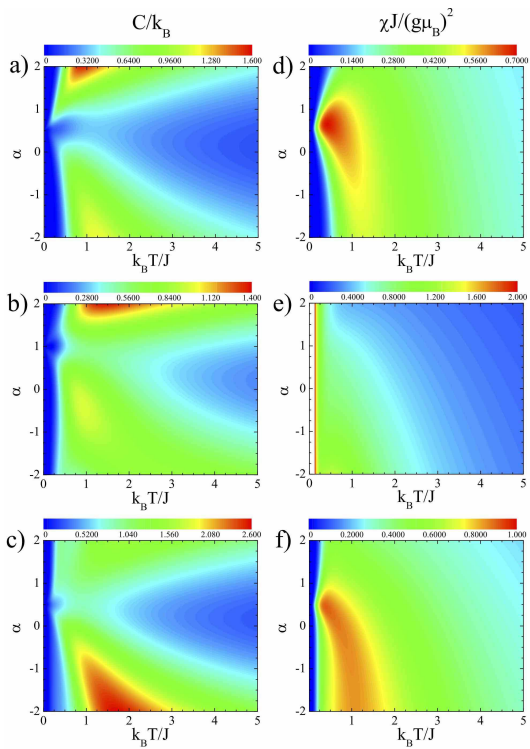}
\vskip -0.2cm
\caption{(color online) Contour plots of the calculated specific heat (a-c) and magnetic susceptibility (d-f) for the  square (a,d), pentagon (b,e), and hexagon (c,f) as function of temperature and $\alpha$.}
\label{c-chi-alpha}
\end{figure}

Figure \ref{c-chi} shows the calculated specific heat and magnetic susceptibility for the equilateral triangle, square, pentagon, and hexagon as function of temperature with $\alpha$=0 (no next-nearest neighbor interactions), while Fig. \ref{c-chi-alpha} show a contour plot of the calculated specific heat and magnetic susceptibility for the square, pentagon, and hexagon as a function of temperature and $\alpha$.

In general, the specific heat demonstrates a distinct enlargement in the Schottky anomaly from $n$ = 3 to 6. Overall, Figure \ref{c-chi}(a) shows the standard specific heat response from an antiferromagnet, where the approximate Schottky peak position is around $J$ = $k_BT$. However, as shown in Fig. \ref{c-chi-alpha}(a-c), when $\alpha$ is added ferromagnetically ($\alpha<0$), the specific heat peak position is shifted to higher temperatures, which gives general method for assessing the coupling strength for the crossover interactions. However, as $\alpha$ increases antiferromagnetically ($\alpha>0$), there is consistently a drop in specific heat and then a shifting in temperature.

In reference to the magnetic susceptibility, Figure \ref{c-chi}(b) shows the standard ferromagnetic and antiferromagnetic responses. Because the equilateral triangle and pentagon are odd numbered clusters, they have non-zero ground states and therefore provide a typical ferromagnetic response by increasing exponentially as the temperature approaches zero. However, the square and hexagon do have spin-0 ground states and show the general peak and reduction to zero as temperature approaches zero.

\section{Inelastic Neutron Scattering Structure Factors}

As molecular magnet systems grow in size, the larger magnetic system can be thought of as a network of the smaller subgeometries in which excitations of the larger system can be described through an analysis of the smaller groups\cite{hara:11}. Therefore, regardless of spin, the inelastic neutron scattering structure factor or intensity for the discrete excitations can be characterized by the subgeometries that are being excited, where the functional form of the intensity is specific to that subgeometry (i.e. a tetramer can breakdown into two coupled dimers). 

The ability to breakdown the larger system into smaller subgeometries is critical to the understanding of nanomagnetic systems. Although, two main complications arise\cite{hara:11}:

\begin{itemize}
\item When there exists a large amount of spin-mixing in the subgeometries (i.e. spin sites are coupled to two or more subgeometries.) This leads to off-diagonal terms in the Hamiltonian that can complicate the determination of the eigenvalues and vectors for the system.

\item When the spin of the system is mixed valence or high spin, where the number of states greatly complicates the Hamiltonian and the individual eigenvalues are difficult to determine. 
\end{itemize}

However, the functional form of the structure factor is not dependent on the spin the of the system and the spin mixing does not effect the individual subgeometry excitations. Therefore, having an understanding of the structure factors produced by specific subgeometries can help in the characterization of larger magnetic systems.

To gain further information about the magnetic structure and interactions, the zero temperature magnetic neutron scattering differential cross section for the inelastic scattering of an incident neutron from a magnetic system in an initial state $|\Psi_i\rangle$\cite{Squ78} can be determined. Considering momentum transfer $\hbar\vec q$ and energy transfer $\hbar\omega$, the differential cross section is proportional to the neutron scattering structure factor tensor
\bd
S_{ba}(\vec q, \omega) = \hskip 3cm
\ed
\be
\int_{-\infty}^{\infty}\! \frac{dt}{2\pi} \
\sum_{\vec x{_i}, \vec x{_j}}
e^{i\vec q \cdot (\vec x_i - \vec x_j )  +i\omega t}
\langle \Psi_i |
\S_b^{\dagger}(\vec x_j, t) \S_a(\vec x_i, 0)
| \Psi_i \rangle \ .
\label{Sab_def0}
\ee
The site sums in Eq.(\ref{Sab_def0}) run over all magnetic ions in one unit cell, and $a,b$ are the spatial indices of the spin operators\cite{hara:05}.

For transitions between discrete energy levels, the time integral gives a trivial delta function
$\delta(\E_f - \E_i - \hbar \omega)$ in the energy transfer, so it is useful to specialize to an ``exclusive structure factor" for the excitation of states within a specific magnetic multiplet (generically $|\Psi_f (\lambda_f)\rangle $)
from the given initial state $|\Psi_i \rangle $,
\be
S_{ba}^{(fi)}(\vec q\, ) =
\sum_{\lambda_f}\
\langle \Psi_i |
V_b^{\dagger}
| \Psi_f (\lambda_f)\rangle \ \langle \Psi_f (\lambda_f)|
V_a
| \Psi_i \rangle  \ ,
\label{Sab_def}
\ee
where the vector $V_a(\vec q\,) $ is a sum of spin operators over all magnetic ions in a unit cell,
\be
V_a = \sum_{{\vec x}_i} {\S}_a(\vec x_i)\;
e^{i\vec q \cdot \vec x_i } \ .
\label{Va_defn}
\ee

\begin{figure}
\includegraphics[width=3.75in]{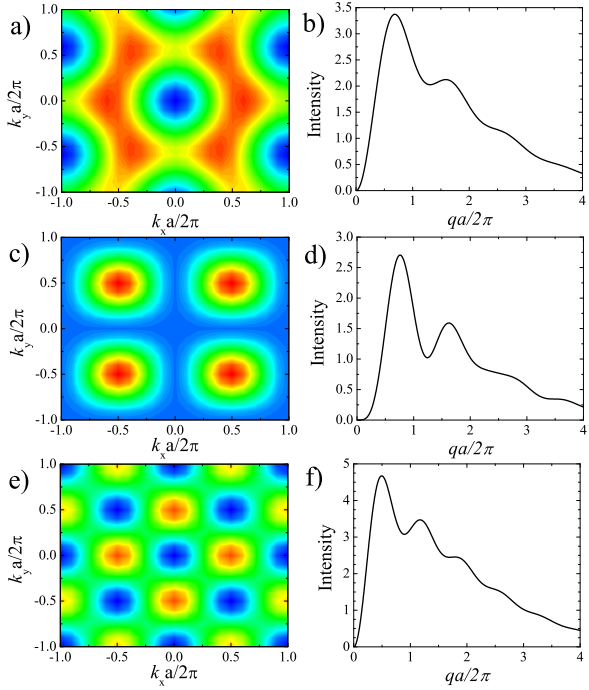}
\vskip -0.2cm
\caption{(color online) The calculated total structure factor (a,c, and e) and powder average intensities (b, d, and f) as a function of $k_x$ and $k_y$ for the spin-1/2 equilateral triangle (a and b) and square ($|0\big>_{I}\rightarrow|1\big>_{I}$ (c-d) and $|0\big>_{I}\rightarrow|1\big>_{II}$ (e-f)).}
\label{t-t-sf}
\end{figure}

Here, we simplify the presentation by quoting the unpolarized result $\langle S_{ba}^{(fi)}(\vec q\, )\rangle$,
obtained by summing over final and averaging over initial polarizations. This unpolarized
$\langle S_{ba}^{(fi)}(\vec q\, )\rangle$ {\it is} $\propto \delta_{ab}$, so it suffices to give the function $S(\vec q\, )$;
\bd
\langle S_{ba}^{(fi)}(\vec q\, )\rangle  =  \delta_{ab}\, S(\vec q\, )
=
\ed
\be
\frac{1}{2\S_i+1}\sum_{\lambda_i,\lambda_f}\
\langle \Psi_i (\lambda_i) |
V_b^{\dagger}
| \Psi_f (\lambda_f)\rangle\
\langle \Psi_f (\lambda_f)|
V_a
| \Psi_i(\lambda_i) \rangle \ .
\label{Strunpoldefn}
\ee
If desired, the general results for polarized scattering can be recovered by reintroducing the appropriate Clebsch-Gordon coefficients in Eq.(\ref{Sab_def}).

The results given above apply to neutron scattering from single crystals. To interpret neutron experiments on powder samples,  we require an orientation average of the unpolarized single-crystal neutron scattering structure factor. Therefore, we define this powder average by
\be
{\bar S}(q) = \int \frac{d\Omega_{\hat q}}{4\pi}\, S(\vec q\, ) \ ,
\label{Strpowavg}
\ee
which is an integration over all angles. In general, the functional form of the single crystal structure factor is a sum of various cosine functions of $\vec q \cdot \vec r_{ij}$, where $\vec r_{ij}$ is $\vec r_{j} - \vec r_{i}$. When integrated over all angles, $\cos(\vec q \cdot \vec r_{23})$ becomes a $0^{th}$ order Bessel function $j_0(q |r_{ij}|)$, where $j_0(x) = \sin(x)/x$. It should noted that structure factor functions presented are not normalized, since the identification of an excitation only requires the functional form.

Evaluating the eigenvalues for the Hamiltonian matrix provides the eigenstates given in Tables \ref{t-t-values}-\ref{h-values}. The eigenstates as a function of $\alpha$ for the square, pentagon, and hexagon are given in Fig. \ref{structures}. Here, it is clear that the ground state for each system changes depending on the value of $\alpha$. Therefore, for consistency, we only consider the case where $\alpha$ = 0, since the functional forms of the transitions will be similar.

\begin{figure}
\includegraphics[width=3.75in]{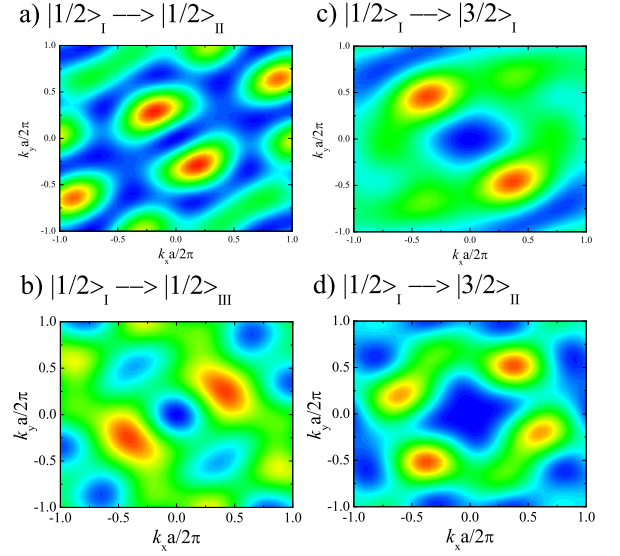}
\vskip -0.2cm
\caption{(color online) The calculated total structure factor intensity for all four as a function of $k_x$ and $k_y$ for the spin-1/2 pentagon.}
\label{p-sf}
\end{figure}

Here, it is important to note that the equilateral triangle and the pentamer have spin-1/2 ground states, while the square and hexagon have spin-0 ground states. This is critical because the spin-0 ground states introduce only singlet ground states, while the spin-1/2 states are doublets. Furthermore, the triangle and pentagon have doubly degenerate doublets. Therefore, the number of total transitions are increased. However, that simply affects the normalization coefficients and not the functional form for each transition. Although, there could exist a combination of multiple transitions as will be shown in both cases. 

\subsection{$S=\frac{1}{2}$ Triangle and Square}

The structure factor for the equilateral triangle was previously reported by Haraldsen et al. in 2005\cite{hara:05}. However, it is included in Table \ref{t-t-values} for consistency and comparison to the larger systems. Figure \ref{t-t-sf}(a) and (b) shows the single crystal and powder average response for an excitation of triangle.
The structure is combination of the trimer excitations plus the added structure factor of the dimer transition from the second ground state. Furthermore, Figure \ref{t-t-sf}(c)-(d) shows the intensity patterns for the square, which has three different functional forms that are all tetramer excitations (see Table \ref{t-t-values}).

Interestingly, the structure factors for the triangle and the square are only excitations to the first excited state. Although, if the ground state degeneracy is broken for the equilateral triangle, then that transition will be that of a dimer ($1-\cos(\vec q \cdot \vec r_{13})$)\cite{hara:05}. The spin-0 transition for the square ($|0\big>_{I}\rightarrow|0\big>_{II}$) has zero intensity because it is forbidden due to geometric selection rule restrictions put on by cluster excitations\cite{hara:11}. 

Furthermore, it is clear to see that while powder average intensities are similar for the equilateral triangle and the square, the single crystal responses are quite different. This points out the constraints on inelastic neutron scattering on powders as opposed to single crystals.

\subsection{$S=\frac{1}{2}$ Pentagon}

As shown in Table \ref{p-values}, the energy eigenstates become very complex for the spin-1/2 pentagon. Here, there is a doubly degenerate ground state as well as multiple doubly degenerate excited states that produce four total excitations. However, due to the degenerate ground state and the multiple excited states, there are a total of 14 total structures for this system. 

\begin{figure}
\includegraphics[width=3.75in]{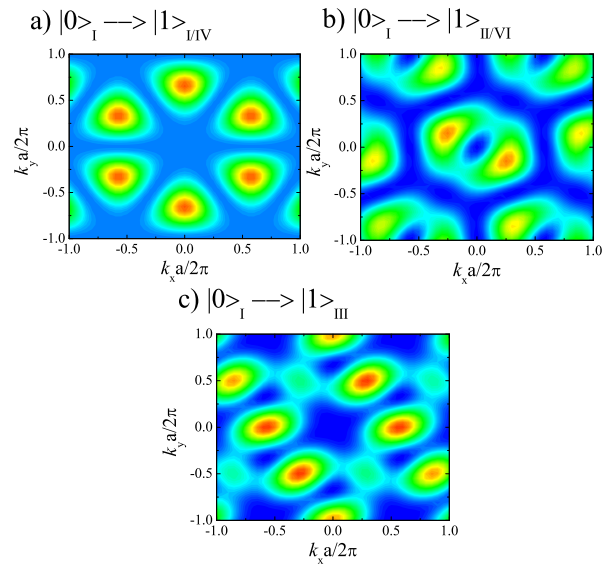}
\vskip -0.2cm
\caption{(color online) The calculated total structure factor intensity for all five transitions as a function of $k_x$ and $k_y$ for the spin-1/2 hexagon.}
\label{h-sf}
\end{figure}

Furthermore, because the pentagon cannot break down into individual subgeometries, there is complex spin mixing of the states. This is even more complicated by the non-zero ground state. Therefore, the eigenfunctions produced by the Hamiltonian matrix are have intricate coefficients that manifests the detailed structure factors shown in Table \ref{p-values}. The total intensity for each transition is a linear combination of the possible structure factors for that excitation from both ground states. 

Figure \ref{p-sf} shows the single crystal structure factor intensity for all four transitions of the pentagon.  Here, the individual intensities presented in Fig. \ref{p-sf} are weighted equally in determining the total structure factor. This demonstrates the overall increase in complexity going from the straightforward equilateral triangle ($n$ = 3) with a doubly degenerate ground state to the pentagon ($n$ = 5). This complexity is due to the spin mixing between the eigenstates.

\subsection{$S=\frac{1}{2}$ Hexagon}

While being a larger system than the pentagon, the spin-1/2 hexagon is not nearly as complex. This is due to the presence of a spin-0 ground state. As shown by the energy eigenvalues in Table \ref{h-values}, the hexagon systems does have mixing between the spins, which is also evident by the interactions in the structure. However, even with spin mixing of states, the calculated structure factor intensities are fairly straightforward (as shown in Table \ref{h-values}. 

As with S =1/2 square, the hexagon does not produce a transition between the spin-0 states. This is due to the subgeometry restrictions to neutron scattering selection rules. This is also evident in the zero intensity transition from $|0\big>_{I}$ to $|1\big>_{V}$. However, the other spin-1 states due produce definite transitions. Figure \ref{h-sf} shows the calculated structure factors as a function of $k_x$ and $k_y$.

Overall, as one increases the number of spin within the clusters, it is relevant to recognize the overall symmetry of the clusters as well as the effective ground state. While the intensity functions are complex, they will be consistent with increasing spin. In systems of $S = 1$ or greater for the same geometries, one should expect that the functional forms will be the same. The overall normalization form will change, but in comparison with experiment, the overall normalization is not always needed.

\section{Conclusions}

In conclusion, the energy eigenstates, magnetic properties, and structure factor intensities for a quantum spin-1/2 equilateral triangle, square, pentagon, and hexagon are determined and compared for $n$ = 3, 4, 5, and 6 spin-1/2 quantum rings. In some cases, the general functional form of the energy states is given. Overall, these systems were modeled using an isotropic Heisenberg Hamiltonian with nearest and next-nearest neighbor interactions. Overall, this study shows the increasing complexity in the energy eigenstates and evolution of the magnetic properties and inelastic neutron scattering structure factors as the quantum rings are increased in size and as cross interactions are enabled. Furthermore, the presence of a complex spin-mixing, multiple ground states, and non-zero ground states greatly complicates the spin Hamiltonian and produces a large complexity in the functional forms of the structure factors.

The goal of this work is to provide insight into the evolution of the spin excitations within these systems for the use in understanding transitions in larger molecular magnet systems. Moreover, this work can be useful in providing insight into the interactions of larger nanoparticles on two-dimensional structures.

\section*{Acknowledgements}

J.T.H would like to thank O. Zaharko and T. Pekerak for useful and insightful discussions. This work was partially supported by Institute for Materials Science at Los Alamos National Laboratory.

\end{document}